\documentclass[12pt]{article}
\usepackage[dvips]{graphicx}
\usepackage[dvips]{graphics}
\usepackage{enumerate}
\usepackage{psfrag}
\newcommand{\newc}{\newcommand}
\newc{\ra}{\rightarrow}
\newc{\lra}{\leftrightarrow}
\newc{\beq}{\begin{equation}}
\newc{\eeq}{\end{equation}}
\newc{\barr}{\begin{eqnarray}}
\newc{\earr}{\end{eqnarray}}
\newc{\texa}{\textstyle}
\newc{\paral}{\parallel}
\newc{\und}{\underline}
\newc{\pars}{\partial}
\newc{\nonu}{\nonumber \\}
\newc{\jump}{\nonumber \\[2.0mm]}
\newc{\rar}{\rightarrow}
\newc{\al}{\alpha}

\begin{document}
\begin{center}
{\Large \bf Exact magnetohydrodynamic equilibria with \\
\vspace{2mm} flow and effects on the Shafranov shift\footnote{  Presented at the 5th Intern. Congress on Industrial and
Applied Mathematics, 7-11 July 2003, Sydney, Australia.}}
 \\ \vspace{2mm}

 {\large G. N. Throumoulopoulos$^\dag$\footnote{Electronic mail: gthroum@cc.uoi.gr},
  G. Poulipoulis$^\dag$, G. Pantis$^\dag$,
H. Tasso$^\star$\footnote{Electronic mail:
henri.tasso@ipp.mpg.de}}
\vspace{1mm}\\
{\large \it $^\dag$ University of Ioannina, \\   Association
Euratom - Hellenic Republic,\\ Physics Department, Section of
Theoretical Physics, \\ GR 451 10 Ioannina, Greece}\\ \vspace{1mm}

{\large \it $^\star$ Max-Planck-Institut f\"{u}r Plasmaphysik,
\newline
Euratom Association,\\ \vspace{1mm}
 D-85748 Garching, Germany}
\end{center}
\vspace{0.1cm}
%
%
\begin{center}
{\large\bf Abstract}
\end{center}

 Exact solutions of the equation governing the
 equilibrium magnetohydrodynamic
 states of an axisymmetric plasma with
incompressible flows of arbitrary direction [H. Tasso and G.N.
Throumoulopoulos, Phys. Pasmas {\bf 5}, 2378 (1998)] are
constructed  for toroidal current density profiles peaked on the
magnetic axis in connection with the ansatz $S=-ku$, where $S=d/d
u\left\lbrack \varrho (d\Phi/du)^2\right\rbrack$  ($k$ is a
parameter, $u$ labels the magnetic surfaces; $\varrho(u)$ and
$\Phi(u)$ are  the density and the electrostatic potential,
respectively). They  pertain to either unbounded plasmas of
astrophysical concern or bounded plasmas of arbitrary aspect
ratio. For $k=0$, a case which includes flows parallel to the
magnetic field, the solutions are expressed in terms of Kummer
functions while for $k\neq 0$ in terms of Airy functions. On the
basis of a tokamak solution with $k\neq 0$ describing a plasma
surrounded by a perfectly conducted boundary of rectangular
cross-section
 it
turns out that the Shafranov shift is a decreasing function which
can vanish for a positive value of  $k$. This value is larger the
smaller the aspect ratio of the configuration.
\newline

{\large\sf Published in Phys. Lett. A {\bf 317}, 463-469 (2003)}

\newpage
\begin{center}
{\large \bf 1.\ \ Introduction}
\end{center}

Flow is a common phenomenon in astrophysical plasmas, in
particular let us mention  here the plasma jets that are observed
to be spat out from certain galactic nuclei and propagate to
supergalactic scales, e.g. Ref. \cite{Eil}. Also, there has been
established in fusion devices that sheared flow can reduce
turbulence either in the edge region (L-H transition) or in the
central region (internal transport barriers) thus resulting in a
reduction of the outward particle and energy transport, e.g. Ref.
\cite{Te}.

In an attempt to contribute to the understanding of the
equilibrium properties of flowing plasmas we considered
cylindrical \cite{ThPa,ThTa}, axisymmetric \cite{TaTh,SiTh},  and
helically symmetric \cite{ThTa1}  steady states with
incompressible flows in the framework of ideal
magnetohydrodynamics (MHD) by including the convective flow term
in the  momentum equation. Also we studied the equilibrium of a
gravitating  plasma with incompressible flow confined by a
point-dipole magnetic field \cite{ThTa00}. For an axisymmetric
magnetically confined plasma the equilibrium satisfies [in
cylindrical coordinates ($z, R, \phi$) and convenient units] the
elliptic differential equation for the poloidal magnetic
flux-function $\psi$,  \beq (1-M_p^2) \Delta^\star \psi -
         \frac{1}{2}(M_p^2)^\prime |\nabla \psi|^2
                     + \frac{1}{2}\left(\frac{X^2}{1-M_p^2}\right)^\prime
+ R^2\left( P_s \right)^\prime + \frac{R^4}{2}\left(\frac{\rho
(\Phi^\prime)^2}{1-M_p^2}\right)^\prime
    = 0,
                            \label{1}
\eeq
along with a Bernoulli relation for the pressure, \beq P=P_s(\psi)
- \varrho\left\lbrack \frac{v^2}{2} - \frac{R^2
(\Phi^\prime)^2}{1-M_p^2}\right\rbrack.
                                  \label{1a}
\eeq
 Here, $P_s(\psi)$,
$\rho(\psi)$ and  $\Phi(\psi)$ are, respectively, the
static-equilibrium pressure, density and electrostatic potential
which remain constant on magnetic surfaces
$\psi(R,z)=\mbox{constant}$; the flux functions $F(\psi)$ and
$X(\psi)$  are  related to the poloidal flow and the toroidal
magnetic field;  $M_p= F^\prime/\sqrt{\varrho }$ is the
Mach-number of the poloidal velocity with respect to the
poloidal-magnetic-field Alfv\'en velocity; $\Delta^\star\equiv
R^2\nabla\cdot(\nabla/R^2)$; the prime denotes differentiation
with respect to $\psi$.
For vanishing flow (\ref{1}) and (\ref{1a}) reduce to the
 Grad-Schl\"uter-Shafranov equation and $P=P_s(\psi)$, respectively.
 Derivation of (\ref{1}) and (\ref{1a}) is
 given in Ref. \cite{TaTh}.  It should be clarified that to simplify (22) of [5], the
     flux-function term $P_s-XF^\prime\Phi^\prime/(1-M^2)$ therein has been replaced by $P_s$ in (1)
     here. Consequently, relation (19) of [5] for the pressure takes the
     form (2) here. Also, to emphasise that a poloidal-velocity Mach
     number is involved, $M$ in [5] is denoted by $M_p$ here.
 

Under the transformation
\beq u(\psi)= \int_0^{\psi}\, [1-M_p^2(g)]^{1/2}\, d\,g,
     \ \    M_p^2<1,
                               \label{2}
\eeq (\ref{1}) reduces
to
\begin{equation}
  \Delta^\star u
+ \frac{1}{2}\frac{d}{du}\left(\frac{X^2}{1-M_p^2}\right) +
R^2\frac{d P_s}{d u}
 + \frac{R^4}{2} \frac{d}{du}
\left\lbrack\varrho \left(\frac{d \Phi}{du}\right)^2\right\rbrack
= 0.
                            \label{3}
\end{equation}
  The  flow contributions here are connected with $M_P^2$ and $S \equiv d/d
u\left\lbrack \varrho \left(d\Phi/du\right)^2\right\rbrack$.
Equation (\ref{3}), free of the nonlinear term
$1/2(M_p^2)^\prime|\nabla\psi|^2$, can be analytically solved by
assigning the $u$-dependence of the  free flux functions $P_s$,
$X$, $F$, $\varrho$, and $\Phi^\prime$. Relation (\ref{1a}) then
determines the pressure.  For $S=$ constant, exact solutions which
extend the well known Solov\'ev one were derived in Ref.
\cite{SiTh}. Owing to the flow and its shear a variety of new
configurations are possible having either one or two stagnation
points in addition to the usual ones with a single magnetic axis.

The aim of the present work is to construct exact solutions of
 (\ref{3}) for  $S\propto u$  and examine their
properties. The solutions  are of the
form (\ref{5}) below in Section 2. This form is advantageous in
that boundary conditions associated with either bounded laboratory
plasmas or unbounded plasmas of astrophysical concern can be
treated in a unified manner by adjusting appropriately the
parameters it contains. As a matter of fact certain of the
solutions we construct constitute extensions of the
Hernegger-Maschke solution of the Grad-Schl\"uter-Shafranov
equation for bounded plasmas \cite{He,Ma} and a recent one for
unbounded plasmas derived
 by Bogoyavlenskij \cite{Bo00}.

The work is organized as follows. Exact solutions of (\ref{3}) for
$S\propto u$ are constructed in Section 2.  The flow impact on the
Shafranov shift is then examined in Section 3 on the basis of a
particular solution  describing a tokamak configuration of
arbitrary aspect ratio being contained within a perfectly
conducting boundary of rectangular cross-section. Section 4
summarizes the Conclusions.
\newpage

\begin{center}
{\large \bf 2.\ \ Exact equilibrium solutions}
\end{center}

 Let us  make the following particular choice for  the free functions:
 \beq \frac{X^2}{1-M_p^2}= X_0^2 + c_1^2 u^2,\ \ P_s=P_{s0}
+ 2 (-1)^\gamma c_2^2  u^2,\ \  \mbox{and} \ \ S\equiv
\left\lbrack \rho(\Phi^\prime)^2\right\rbrack^\prime = -64 c_3
c_2^3 u.
                                                       \label{4}
\eeq
under which  (\ref{3}) becomes  linear. 
Here, the factor $(-1)^\gamma$ has been introduced in order
to make comparison with solutions existing in the literature
convenient, and  $X_0$, $P_{s0}$, $c_1$, $c_2$, $c_3$ are
constants.
 Equation (\ref{3}) then takes the form
 \beq \Delta^\star u +c_1^2 u  + 2(-1)^\gamma c_2x u -8c_2c_3x^2  u
 =0,                                                    \label{4a}
 \eeq
 where $x=2c_2 R^2$.
 We  pursue
separable solutions of the form
 \beq u(R,z)=x^n
P(x) T(z)\exp\left(-\frac{\gamma x}{2}\right).
                                                         \label{5}
\eeq This is appropriate for considering various equilibrium
configurations in connection with different boundary conditions.
In particular, the term $x^n$ which makes $u$ to vanish on the
axis of symmetry is associated with either compact tori or
spherical tokamak configurations. Plasmas surrounded by a fixed
perfectly conducting boundary can be considered by setting
$\gamma=0$. Unbounded plasmas are connected with  $\gamma=1$,
 the exponential term
guaranteeing smooth behaviour at large distances.
It is noted that for static equilibria ($M_p=c_3=0$) the cases $n=
\gamma=0$ and ($n=0, \gamma=1$) were considered in Refs.
\cite{He,Ma}, and \cite{Bo00}, respectively.

 With the use of (\ref{5})
equation (\ref{4a}) leads to the following  differential equations
for $P(x)$ and $T(z)$: \beq T^{\prime\prime} + \omega^2 T=0
                                                       \label{6}
\eeq
and
\barr xP^{\prime\prime}+ (2n - \gamma x) P^\prime & &
\nonu  +\left\lbrack \frac{4n(n-1) - 4\gamma n x + \gamma^2 x^2}{4
x}
   +\frac{(-1)^\gamma x}{4}-\tau - c_3x^2 \right\rbrack P =0, &
   &
                                                        \label{7}
\earr
 where $\omega^2=$ constant and $\tau=(\omega^2 - c_1^2)/8c_2$.
 Therefore,
 \beq
 T(z)= A\sin(\omega z) + B \cos(\omega z)
                                                   \label{7a}
 \eeq
and  the problem  additionally requires to solving  (\ref{7}).
Solutions of   (\ref{7}) associated with  different  boundary
conditions  will  be constructed in the subsequent subsections. It
is noted here that all the solutions  hold for arbitrary poloidal
Mach numbers, viz. the dependence of $M_p^2(u)$ on $u$ remains
free.

\begin{center}
 {\large \em 2.1\ \ Unbounded
plasmas ($\gamma=1$) and $c_3=0$}
 \end{center}

 It is first noted that the case $c_3=0$ includes flows parallel to the
 magnetic field  for $\Phi^\prime=0$, viz.  when the  electric field
 vanishes.
 For $c_3=0$,  (\ref{1})
becomes identical in form with the Grad-Schl\"uter-Shafranov
equation and
  (\ref{7}) reduces to
\beq x^2 P^{\prime\prime} +x(2n-x) P^\prime + \left \lbrack n(n-1)
-(\tau +n)x\right\rbrack P =0.
                                                             \label{9}
\eeq
 Equation (\ref{9}) can be solved by  the following
procedure.  The substitution
 $P=x^k W(x)$, where
 $k$ is a root of the quadratic equation
\beq
 k^2 + (2n -1)k + n(n-1)=0,
                                 \label{10}
\eeq
 leads to the following equation for $W(x)$:
 \beq
xW^{\prime\prime} + [2(k+n)-x]W^\prime  -(k+\tau + n)W =0.
                                                        \label{11}
\eeq The solutions of this equation can be expressed in terms of
 Kummer or confluent hypergeometric functions  (\cite{Abr}, p. 503; \cite{PoZa}, p. 137).
In particular,  for the two roots of  (\ref{10}) we have:

\noindent {\em 2.1.a\ \  $k_1=-n$}

The solution of  (\ref{11}) is
$$W(x)= x\left\lbrack D_1 M(\tau +1,2,x) +
D_2 U(\tau +1, 2, x)\right\rbrack,$$
 where $M$ and $U$ are the  Kummer
functions of first and second kind, respectively, and $D_1$, $D_2$
are constants. Consequently, the solution of the original equation
(\ref{3}) is written in the form
 \barr
u(x,z)&=& x\left\lbrack D_1 M(\tau +1,2,x) + D_2 U(\tau +1, 2,
x)\right\rbrack  \exp{\left(-x/2\right)} \nonumber \\
& & \left\lbrack A \sin(\omega z) + B \cos(\omega z)\right\rbrack.
                                                     \label{12}
\earr
 For special values of $\tau$ and $n$ the Kummer functions reduce to simpler
 classical functions or polynomials (see Ref. \cite{Abr}, p. 509, table 13.6) ;
  in particular,  for $\tau=-m$,
 where $m $ is a non-negative integer, and $n=0$ they reduce to Laguerre polynomials.
Solutions
 of this kind for
 static equilibria ($M_p=0$) were  derived in Ref. \cite{Bo00}
 and   employed to model
astrophysical jets and solar prominences. It may also be noted
here that a continuation of this study and related studies  were
reported
in Refs. \cite{Bo00a,Bo01} and \cite{Nu}.\\

\noindent {\em 2.1.b\ \  $k_2=-(n + 1)$}

The  solutions of (\ref{11}) and (\ref{3}), respectively, read
\beq W(x)=  x^3\left\lbrack D_1 M(\tau +2,4,x) + D_2 U(\tau +2, 4,
x)\right\rbrack, \eeq and\barr u(x,z)&= &x^2[D_1 M(\tau +2,4,x) +
D_2
U(\tau +2, 4, x)]\exp{\left(-x/2\right)} \nonumber \\
& & \left\lbrack A \sin(\omega z) + B \cos(\omega z)\right\rbrack.
                                                     \label{13}
\earr

\begin{center}
\noindent {\large \em 2.2\ \  Bounded plasmas ($\gamma=0$)\ and
$c_3= 0$
                     }
\end{center}

Equation (\ref{7}) then reduces to
 \beq
 x^2P^{\prime\prime} +2n x
P^{\prime} + \left\lbrack n(n-1) -\tau x -\frac{1}{4} x^2
\right\rbrack =0.
                                                   \label{14}
\eeq This equation can be solved by a procedure similar to that in
Sec. 2.1, i.e the substitution $P=x^k W(x)$, where  $k$ is again a
root of  (\ref{10}),  leads to \beq xW^{\prime\prime} + 2(n+k)
W^\prime -
 \left(\frac{1}{4}x+\tau\right) W =0.
                                                   \label{14a}
 \eeq
 Equation (\ref{14a}) has solutions of the form $W(x)=\exp(x/2)g(x)$.
For the two roots of (\ref{10}), $k_1=-n$ and $k_2=-(n+1)$
respectively,
 the solutions  are
 $$ W(x)=\exp(x/2)
\left\lbrack D_1 M(-\tau,0, -x) + D_2 U(-\tau , 0,
-x)\right\rbrack,$$ and
$$ W(x)=\exp(x/2)
 \left\lbrack D_1 M(-(1+\tau),-2, -x) + D_2 U(-(1+\tau) ,
-2, -x)\right\rbrack.$$ The respective solutions for $u(x,z)$ are
\beq u(x,z)= \exp(x/2)\left\lbrack D_1 M(-\tau,0, -x) + D_2
U(-\tau , 0, -x)\right\rbrack T(z)
                                                     \label{15}
\eeq and \beq u(x,z)= \exp(x/2)x^{-1} \left\lbrack D_1
M(-(\tau+1),-2, -x) + D_2 U(-(\tau +1), -2, -x)\right\rbrack T(z)
                                                     \label{16}
\eeq with $T(z)$ as given by (\ref{7a}).

As in Sec. 2.1, for special values of $\tau$ and $n$ the solutions
of  (\ref{14}) can be expressed in terms of simpler classical
functions or polynomials, e.g. for $n=0$  they are expressed in
terms of  Coulomb wave functions.
 In this case respective static solutions ($M_p=0$) were
constructed in Refs. \cite{He} and \cite{Ma}.

\begin{center}
 {\large \em 2.3\ \ Bounded plasmas ($\gamma=0$) and
$c_3\neq 0$}
 \end{center}

 Changing independent variable by
 $$\eta\equiv c_3^{-2/3}\left(c_3 x -\frac{1}{4}\right),$$
 (\ref{7}) for $n=\tau =0$ is transformed into the Airy equation:
\beq \frac{d^2 P(\eta)}{d \eta^2}- \eta P(\eta)=0.
                                                   \label{17}
\eeq The general solution of  (\ref{17}) is \beq P(\eta)=
d_2\left\lbrack B_i(\eta) + d_1 A_i(\eta)\right\rbrack
                                                   \label{18}
\eeq where $A_i$ and $B_i$  are the Airy functions of first and
second kind, respectively, and $d_1$ and $d_2$ are constants. An
equilibrium symmetric with respect to the  mid-plane $z=0$ then is
described by \beq u= d_2\left\lbrack B_i(\eta) + d_1
A_i(\eta)\right\rbrack\cos(c_1 z).
                                                   \label{19}
\eeq For $c_3=0$,
(\ref{19}) takes the simpler form
$$
u= d_2\left\lbrack \sin(\frac{x}{2}) + d_1
\cos(\frac{x}{2})\right\rbrack \cos(c_1 z).
$$
In next section solution (\ref{19}) will be employed to evaluate
the flow impact on the Shafranov shift.
\newpage
\begin{center}
{\large \bf 3.\ \ Flow effects on the Shafranov shift}
\end{center}

 In connection with  solution
(\ref{19}) we now specify the plasma container to have rectangular
cross-section of dimensions $a$ and $b$ and its geometric center
to be located at $R_0$ (Fig. 1). Introducing the dimensionless
quantities $\xi=R/R_0$, $\zeta=z/R_0$, $\lambda=c_1 R_0$, $C=c_2
R_0^2$, and $$H=c_3^{-2/3}\left(2c_3 C\xi^2 - \frac{1}{4}\right)$$
(note that $c_3$ is dimensionless), we  require that $u$ vanishes
on the plasma boundary, viz.
$$u(H=H_\pm)= u(\zeta=\pm
\frac{a}{R_0})=0,$$ where
$$ H_\pm= c_3^{-2/3}\left(2c_3 {C}\xi_\pm^2-\frac{1}{4}\right)\ \  \mbox{and}\ \ \xi_{\pm}\equiv
1\pm \frac{b}{R_0}.$$  This requirement yields the eigenvalues
$$\lambda_l=\frac{R_0}{a}\left
(l\pi+\frac{\pi}{2}\right),\ \ l=0, 1,
 2,\ldots; \ \ {C} = {C}_k,\ \ k= 1,
2,\ldots $$ where ${C}_k$ can be  determined by the equations
$$\frac{B_i(H_k^{+})}{A_i(H_k^{+})}=\frac{B_i(H_k^{-})}{A_i(H_k^{-})}=-D_k,\
\ H_k^{\pm}=c^{-2/3}\left(2c_3 C_k \xi_\pm^2-\frac{1}{4}\right).
$$ The corresponding eigenfunctions normalized to a reference
value $u_c$ are given by $$
\tilde{u}_{kl}=\frac{u_{kl}}{u_c}=\left\lbrack B_i(H_k(\xi) + D_k
A_i(H_k(\xi)) \right\rbrack \cos (\lambda_l \zeta).$$
 The simplest  eigenfunction corresponding to ($k=1$, $l=0$),
\beq
 \tilde{u}_{10}=\frac{u_{10}(\xi,\zeta)}{u_a(\xi=\xi_a,\zeta=0)}=\frac{\left\lbrack B_i(H_1(\xi)) + D_1 A_i(H_1(\xi))\right \rbrack
                 \cos(\lambda_0 \zeta)}{ B_i(H_1(\xi_a)) + D_1 A_i(H_1(\xi_a))}
,
                                      \label{20}
 \eeq
describes a configuration with a single magnetic axis located on
$\zeta_a=0$ and  $\xi=\xi_a=1+\Delta \xi$  with $\xi$ satisfying
the equation \beq \frac{d B_i(H_1(\xi))}{d \xi} + D_1 \frac{d
A_i(H_1(\xi))}{d \xi}=0.
                                       \label{21}
\eeq
 Here, $\Delta \xi$ is the Shafranov shift.   Contours of constant $u$ and the Shafranov shift in connection
with (\ref{20}) are illustrated in Fig. 2.

 For $c_3=0$ for which the functions $A_i$ and
$B_i$ reduce to $\cos$ and $\sin$, respectively, the quantities
$C_k$, $D_k$ and $\tilde{u}_{10}$ take the simpler forms:
${C}_k=R_0 k\pi/(4 b)$, $D_k= -\tan({C}_k \xi_{+}^2)=-\tan({C}_k
\xi_{-}^2)$ and
 \beq
 \tilde{u}_{10}=\frac{u_{10}}{u_c}=\left\lbrack \sin({C}_1 \xi^2)+ D_1
 \cos({C}_1 \xi^2)\right \rbrack
                 \cos(\lambda_0 \zeta).
                                      \label{22}
 \eeq
The magnetic axis of the configuration described by (\ref{22}) is
located at ($\zeta=0, \xi=1+\sqrt{1+b^2/R_0^2}$).

The toroidal current density is  given by
$$
j_\phi=\left. \frac{\Delta^\star\psi}{R}=
 \frac{1}{R\left(1-M_p^2\right)^{1/2}}\left\lbrack \Delta^\star u +
 \frac{|\nabla u|^2}{2\left(1-M_p^2\right)}\frac{d M_p^2}{du}\right\rbrack
\right|_{\mbox{\large  u=u}_{10}}.
$$
 For Mach numbers of the form $M_p^2= u^m$, where $m>1$ the profile of
$j_\phi$ is peaked on the magnetic axis and vanishes on the
boundary.

 On the basis of
equation (\ref{21}), $\Delta \xi$ has been determined numerically
as a function of the flow parameter $c_3$ (Fig. 3). It is recalled
that  $c_3$ is related to the density and the electric field and
their variation perpendicular to the magnetic surfaces by
(\ref{4}).  As can be seen in Fig. 3, $\Delta \xi$ is a decreasing
function of $c_3$ which goes down to zero sharply as $c_3$
approaches a positive value, this value being larger the smaller
is the aspect ration $R_0/b$ of the configuration. This result is
independent of the plasma elongation $a/b$. It is finally noted
that suppression of the Shafranov shift by a properly shaped
toroidal rotation profile was reported in Ref. \cite{IlPo}.

\begin{center}
{\large \bf 6.\ \ Conclusions}
\end{center}

 We have constructed  exact solutions of the equation describing the MHD equilibrium states of an
axisymmetric magnetically confined plasma with incompressible
flows [Eq. (\ref{3})] for $S=-ku$, where $S=d/d u\left\lbrack
\varrho (d\Phi/du)^2\right\rbrack$, corresponding to flows of
arbitrary direction.
 The
solutions are  based on  the form (\ref{5}) which is convenient
for applying boundary conditions associated with either unbounded
plasmas or bounded ones.  For $k=0$ the solutions are expressed in
terms of Kummer functions [(Eqs. (\ref{12}) and (\ref{13}) for
unbounded plasmas; (\ref{15}) and (\ref{16}) for bounded ones]
while for $k\neq 0$ they are expressed in terms of Airy functions
[Eq. (\ref{19})].

Solution (\ref{19}) has then been employed  to study  a tokamak
configuration of arbitrary aspect ratio being contained within a
perfectly conducting boundary of rectangular cross-section and
toroidal current density profile which can be peaked   on the
magnetic axis and vanish on the boundary. In this case it turns
out  that the Shafranov shift is a decreasing function of $k$
which can vanish for a positive value $k=k_c$, with $k_c$ being
larger the smaller the aspect ratio of the configuration is. These
results demonstrate that the shape of the density and the electric
field profiles associated with flow and their variation
perpendicular to the magnetic surfaces can result in a strong
variation of the Shafranov shift.

\begin{center}
 {\large\bf Acknowledgements}
\end{center}

Part of this work was conducted during a visit of one of the
authors (GNT) to the Max-Planck Institut  f\"ur Plasmaphysik,
Garching. The hospitality of that Institute is greatly
appreciated.

The present work was performed under the Contract of Association
ERB 5005 CT 99 0100 between the European Atomic Energy Community
and the Hellenic Republic.

\newpage
\begin{center}
{\large \bf  Figure captions}
\end{center}

\noindent Fig. 1 \ Boundary with rectangular cross-section in
connection with the  equilibrium solution (\ref{18}).
\vspace{0.5cm}

\noindent Fig. 2. \  The contours illustrate  magnetic surface cross-sections 
in connection with
solution (\ref{20})  for two different values of the flow parameter $c_3$
which is related to the density and
electric field and their variations perpendicular to the magnetic
surfaces. The positions $\xi_0=1$ and $\xi_a$ refer to the geometric center and the 
magnetic axis, respectively. For $c_3=0.025$  the Shafranov shift $\Delta \xi=(R_a - R_0)/R_0=\xi_a-\xi_0$ in Fig. 2b is drastically
reduced in comparison with that for $c_3=0.005$ in Fig. 2a. \vspace{0.5cm}

\noindent Fig. 3. \ The Shafranov shift $\Delta \xi=(R_a-R_0)/R_0$
versus the flow parameter $c_3$ for various values of the aspect
ratio $R_0/b$. $\Delta \xi$ is determined numerically on the basis
of (\ref{21}) in connection with the equilibrium solution
(\ref{20}). 
\newpage
\begin{figure}[!h]
\begin{center}
\includegraphics[scale=1]{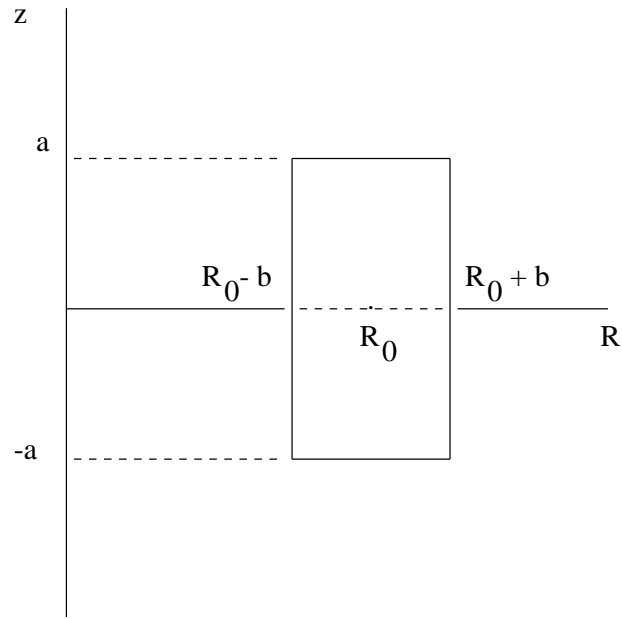}

\caption{Boundary with rectangular cross-section in connection
with the  equilibrium solution (\ref{18}).}
\end{center}
\end{figure}
\begin{figure}[!h]
\begin{center}
\psfrag{z}{$\zeta$ \hspace{3cm} {\huge \cal 2a} \hspace{3cm}
$c_3=0.005$} 
\psfrag{R}{$\xi $} \psfrag{R0}{$\xi_0$}
\psfrag{Ra}{$\xi_a$}
\includegraphics[scale=1]{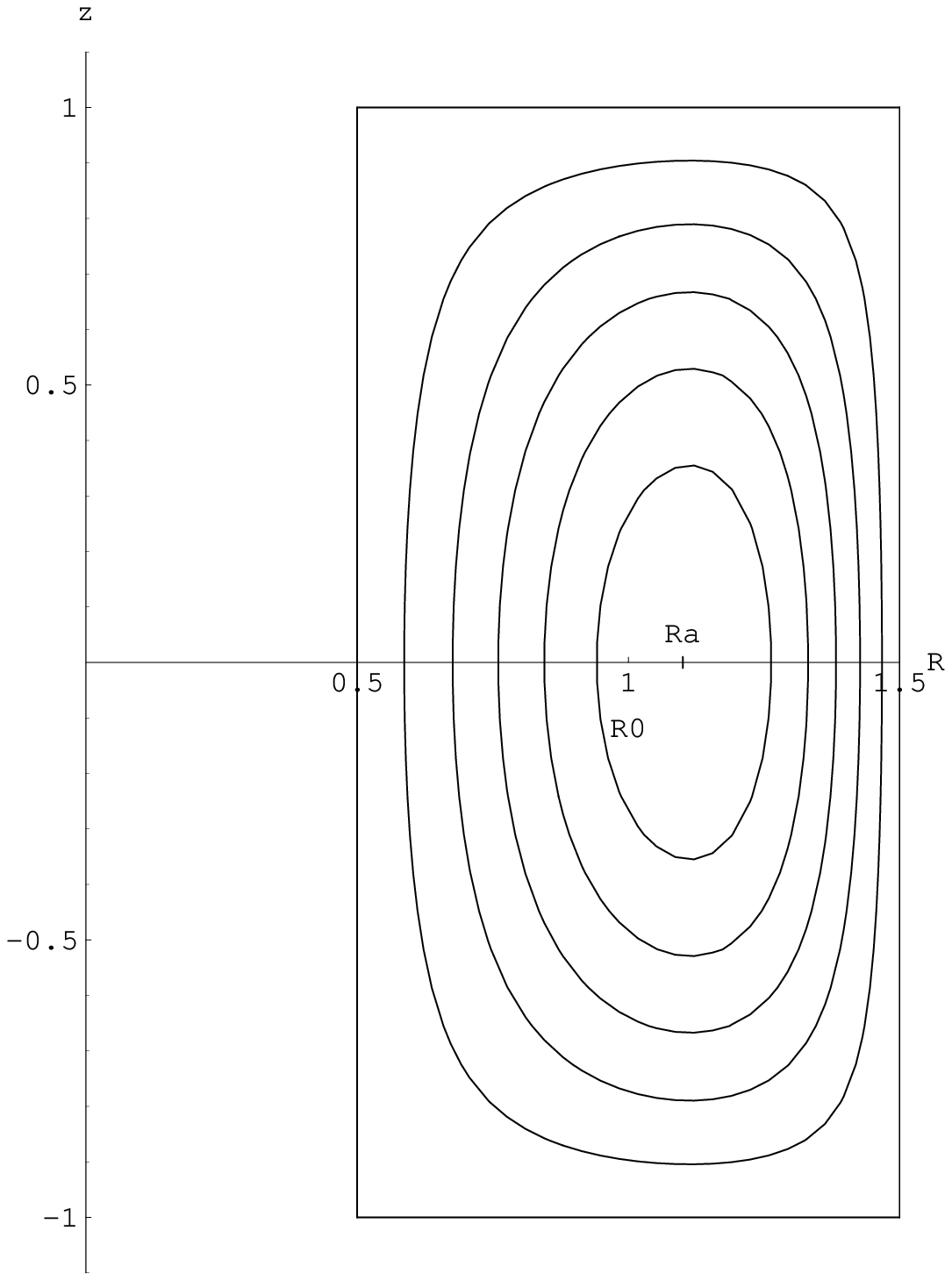}
\end{center}
\end{figure}
\begin{figure}[!h]
\begin{center}
\psfrag{z}{$\zeta$ \hspace{3cm} {\huge \cal 2b} \hspace{3cm}   $c_3$=0.025} 
\psfrag{R}{$\xi $} \psfrag{R0}{$\xi_0$}
\psfrag{Ra}{$\xi_a$}
\includegraphics[scale=1]{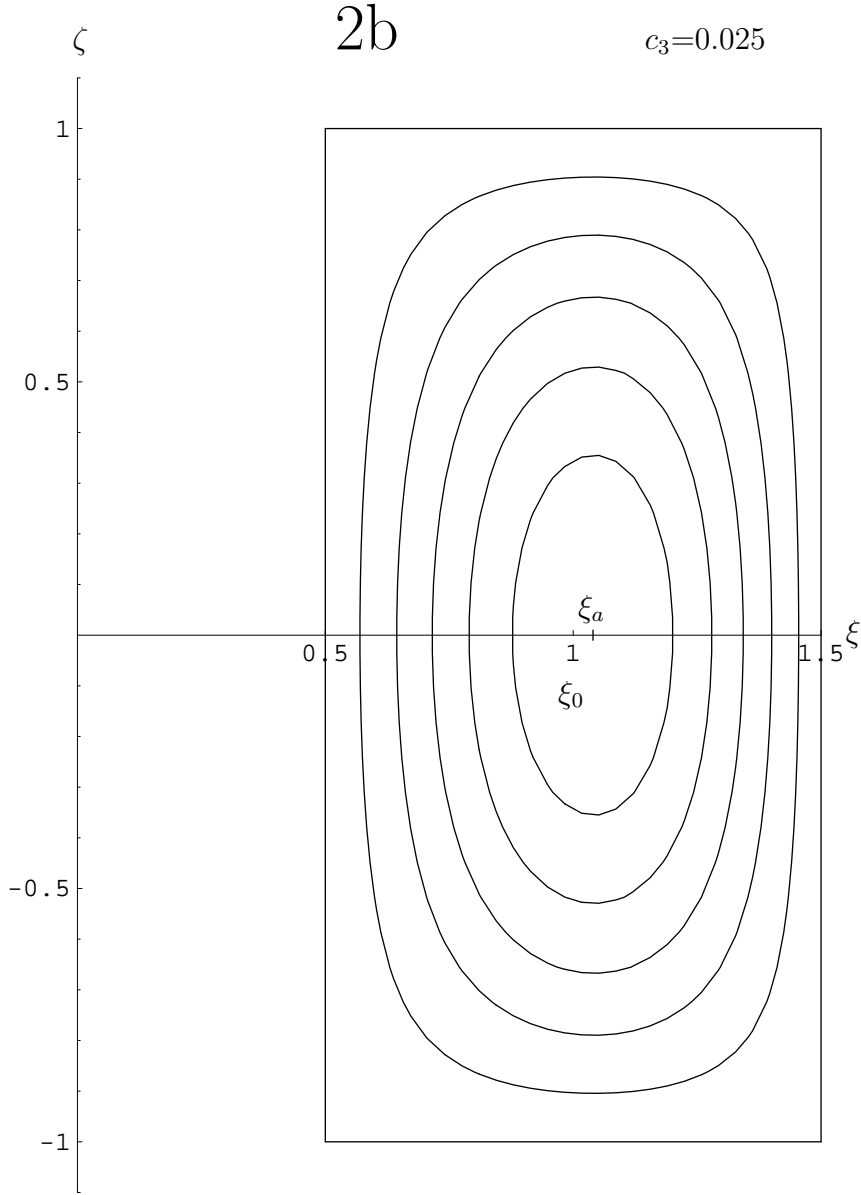}
\caption{The contours illustrate  magnetic surface cross-sections 
in connection with
solution (\ref{20})  for two different values of the flow parameter $c_3$
which is related to the density and
electric field and their variations perpendicular to the magnetic
surfaces. The positions $\xi_0=1$ and $\xi_a$ refer to the geometric center and the 
magnetic axis, respectively. For $c_3=0.025$  the Shafranov shift $\Delta \xi=(R_a - R_0)/R_0=\xi_a-\xi_0$ in Fig. 2b is drastically
reduced in comparison with that for $c_3=0.005$ in Fig. 2a.   }
\end{center}
\end{figure}

%
\begin{figure}[!h]
\begin{center}
\psfrag{a}{$c_3$} \psfrag{b}{$\Delta \xi $} \psfrag{t1}{$R_0/b=1$}
\psfrag{t2}{$R_0/b=2$} \psfrag{t3}{$R_0/b=4$ }
\includegraphics[scale=1]{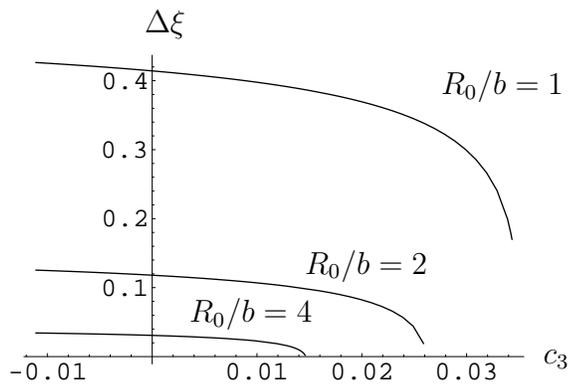}

\caption{  The Shafranov shift $\Delta \xi=(R_a-R_0)/R_0$
versus the flow parameter $c_3$ for various values of the aspect
ratio $R_0/b$. $\Delta \xi$ is determined numerically on the basis
of (\ref{21}) in connection with the equilibrium solution
(\ref{20}).}
\end{center}
\end{figure}
\end{document}